\documentclass[aps,prl,amsmath,amsfonts,showpacs]{revtex4}

\begin{document}

\title{Modern foundations for thermodynamics and the stringy limit of black hole equilibria}

\author{Bernard~S~Kay\footnote{Electronic address: bernard.kay@york.ac.uk}}

\affiliation{Department of Mathematics, University of York, York YO10 5DD, U.K.}

\begin{abstract}
We recall existing string theory work towards an understanding of black hole entropy and we argue that it is incomplete as it stands but we put forward a modified version, based on the author's earlier {\it matter-gravity entanglement hypothesis}, which, we claim, gives a more satisfactory understanding and also a resolution to the Information Loss Puzzle.  This hypothesis pictures a black hole equilibrium state as an, overall pure, state, with given (approximate) energy, consisting of a black hole with its  (mostly matter) atmosphere in a box and identifies the black hole's entropy with the pure state's matter-gravity entanglement entropy.  We assume this equilibrium goes over, at weak string-coupling, to a pure state with similar energy consisting of a long string with a stringy atmosphere and that the matter-gravity entanglement entropy goes over to the entanglement entropy between (approximately) the long string and the stringy atmosphere.  We also recall recent work (in a non-gravitational context) towards modern foundations for thermodynamics, where, in place of a total microcanonical ensemble, one assumes that a total system, consisting of a small (sub)system and an energy bath, is in a (random) pure state with energy in a given narrow range and shows that the small subsystem will then find itself to be in a thermal (Gibbs) state.  We present a new set of formulae, obtained by the author in a companion paper, which generalize the setting of that work to cases where the system and energy bath are of comparable size.  We apply these formulae to a simple model for our string equilibrium where the densities of states of the long string (replacing our energy bath) and stringy atmosphere (replacing our system) both grow exponentially.  We find, for our picture of black hole equilibrium, a temperature of the order of the Hawking temperature and an entropy of the order of the Hawking entropy thus adding to the evidence for the viablity of our matter-gravity entanglement hypothesis and of our picture of black-hole equilibrium states.
\end{abstract}


\maketitle

Thanks to the work \cite{HawkingEvap,HawkingBBBH,GibHawkEuc} of Hawking in the 1970s (see also \cite{Hawking:1982dh}) we know that thermodynamical equilibrium states are in principle possible which consist of a black hole surrounded by a (mainly \cite{KayAby}) matter atmosphere -- all enclosed in a spherical box.  Such states have a temperature, $T$, related to the surface gravity, $\kappa$ ($=1/4G{\cal M}$ for a Schwarzschild black hole of mass $\cal M$) of the black hole by the Hawking formula
$kT=\kappa/2\pi$ ($k$ denotes Boltzmann's constant, $G$ Newton's constant and we set $\hbar=c=1$) and, as anticipated by Bekenstein \cite{Bekenstein:1973ur}, the entropy, $S$, of such a state is proportional to the area, $A$, of the black hole's event horizon where the constant of proportionality takes the Hawking value of $k/4G$.

It is traditionally assumed that the correct mathematical description of such an equilibrium state is as a thermal state in the sense of Gibbs, i.e.\ a (non-pure) density operator of form $\rho=Z^{-1}\exp(-H/kT)$ where $H$ is the total matter-gravity Hamiltonian and $Z$ the partition function, ${\mathrm {tr}}\exp(-H/kT)$.  (A microcanonical state has also been considered \cite{Hawking:1976de} but this is also a non-pure state.)  It is also traditionally assumed that the physical entropy, $S$, of such an equilibrium state should be identified with its von Neumann entropy, $-k\,{\mathrm {tr}}\rho\log\rho$.  However these assumptions lead to puzzles, in particular, to the Information Loss Puzzle \cite{HawkingInfo} which we take here to be the following puzzle:
In the associated, non-equilibrium, process where a (say, spherical) black hole of mass $\cal M$ is formed by stellar collapse and subsequently evaporates, the actual physical entropy shortly after collapse is also believed \cite{HawkingEvap} to be close to the non-zero value $kA/4G$ ($A$ having the Schwarzschild value $16\pi (G{\cal M})^2$) and is expected to increase yet further as the black hole Hawking-evaporates.   Yet if we assume there to be an underlying microscopic description of this quantum gravitational process consistent with a conventional, unitary, formulation of quantum mechanics, in which the state of the total (closed) system before the collapse is (as usually assumed in quantum mechanics) a pure state, then the von Neumann entropy of the total state must start out zero and, being a unitary invariant, remain zero for all time.

Stated in this way, it is apparent that the Information Loss Puzzle is just a special case of the more general, and longstanding, puzzle associated with the Second Law of Thermodynamics, which we take here to be the statement that the entropy of {\it any} closed system increases with time -- the puzzle being that, if the initial state of a closed system is pure and evolves unitarily and if its physical entropy is to be identified with its von Neumann entropy, then one is forced to conclude, in contradiction with the Second Law, that the only value that the entropy can ever take is zero!  (We reject the usual resolutions in terms of coarse-graining as unacceptably subjective.)  In 1998, I proposed \cite{Kay1,Kay2,KayAbyaneh} a resolution to this Second Law Puzzle, and {\it a fortiori} to the Information Loss Puzzle, based on the hypothesis that the physical entropy of a closed system should be identified, not with the von Neumann entropy of its total state but rather with its matter-gravity entanglement entropy.  (Obviously, this proposal requires that we model closed systems as quantum gravitational systems.)  There is then no contradiction between entropy increase (i.e.\ information loss) and unitarity and, indeed, plausibly, if the initial entropy (defined in this way) is low, it will increase monotonically as time increases.

This {\it matter-gravity entanglement hypothesis} entails a radically unconventional description of the equilibrium states of black holes as pure states of matter-gravity, entangled in just such a way that both the matter subsystem and the gravity subsystem are each (approximately) thermal at the Hawking temperature.  Aside from offering a resolution to the Information Loss Puzzle, it also entails that the entropies of the reduced density operators of matter alone and of gravity alone (both necessarily \cite{Kay1,Kaythermality} equal to the matter-gravity entanglement entropy) are both equal to the Hawking entropy, thus offering an explanation \cite{KayAby} for the otherwise unexplained coincidence (Mukohyama and Israel \cite{MukohyamaIsrael} call this coincidence their `correspondence principle') that the entropy obtained from the Gibbons-Hawking action integral \cite{Gibbons:1976ue} for the Euclidean Schwarzschild solution for matterless gravity is equal to $kA/4G$ which is also, as argued by 't Hooft \cite{tHooftBrick} and Mukohyama and Israel \cite{MukohyamaIsrael} and others, equal to the entropy of the thermal atmosphere of an equilibrium black hole in a theory involving matter and gravity even when there are (a large number of) matter fields in the theory.

The full statement \cite{Kay1,Kay2,KayAbyaneh} of our matter-gravity entanglement hypothesis includes, besides this definition of entropy, the closely interrelated postulate that the physically relevant density operator of a closed system, for the purposes of the interpretation of quantum mechanics, is not the total (pure) density operator  $|\Psi\rangle\langle\Psi|$ where $\Psi$ is the total (pure) state vector of the closed system but rather the reduced density operator of matter (i.e.\ the partial trace of $|\Psi\rangle\langle\Psi|$ over gravity).  If this starts out close to pure, it seems reasonable to expect it to get more and more mixed as time increases -- a measure of its degree of mixedness being, in fact, the entropy as we defined it above.

While radically unconventional in the context of quantum black holes, our matter-gravity entanglement hypothesis appears to be in harmony with more recent work (see \cite{GoldsteinLebowitzetal,Popescuetal} and references therein) in a non-gravitational context on the foundations of thermodynamics.  In this work, a new, alternative, explanation is given to the traditional explanation of how small open systems get to be in thermal states.   In both explanations one supposes given a total closed system (we shall from now on, to avoid ambiguous usages of the word `system', use the word `totem' for this) consisting of a small (open) system, S, and an (open) energy bath, B, weakly coupled to one another.  In the traditional explanation, the thermality of S is seen as a consequence of the assumption that the totem is in a microcanonical state.  In the new alternative explanation, which we shall refer to as the `modern' explanation, one assumes the totem to be in a pure state and yet, as shown in \cite{GoldsteinLebowitzetal}, finds that, `typically', the same thermal behaviour of the small open subsystem still results.

Motivated by the problem of understanding the origin of black-hole entropy, in a companion paper \cite{Kaythermality}, we ask what conditions are required for similar scenarios to both the above `traditional' (total-microcanonical-state-) and `modern' (total pure-state-) scenarios for them still to lead to the (approximate) thermality of an open system, when in contact with an energy bath, even when the open system fails to be small but is rather of comparable size to the energy bath.  We next briefly summarize the formalism and results. We denote the density of states of the system, S, by $\sigma_{\mathrm S}(\epsilon)$ and of the energy bath, B, by $\sigma_{\mathrm B}(\epsilon)$.  In both the traditional and modern scenarios, we are interested in the subspace, ${\cal H}_M$, of the totem Hilbert space, spanned by totem energy eigenstates with energies, $\epsilon$, in the range $[E, E+\Delta]$ where $\Delta$ is an energy-increment which is small, yet large enough for the dimension, $M$, of ${\cal H}_M$ to be very large.  We will assume that both $\sigma_{\mathrm S}$ and $\sigma_{\mathrm B}$ are positively supported and monotonically increasing and, for convenience (see again \cite{Kaythermality}) that $E$ is a multiple of $\Delta$ and that both $\sigma_{\mathrm S}$ and $\sigma_{\mathrm B}$ arise from evenly spaced energy levels, 
$\epsilon=0, \Delta, 2\Delta, \dots$ for the same $\Delta$ with degeneracies $n_{\mathrm S}(\epsilon)$
and $n_{\mathrm B}(\epsilon)$ so that $\sigma_{\mathrm S}(\epsilon)=n_{\mathrm S}(\epsilon)/\Delta$ and
$\sigma_{\mathrm B}(\epsilon)=n_{\mathrm B}(\epsilon)/\Delta$.  The total microcanonical density operator,
$\rho_{\mathrm{microc}}$, for totem energies in the range $[E, E+\Delta]$  will then take the form
\begin{equation}
\label{microsysbath}
\rho_{\mathrm{microc}}=
M^{-1}\sum_{\epsilon_{\mathrm{S}}}\sum_{\epsilon_{\mathrm{B}}}\sum_i\sum_j
|\epsilon_{\mathrm{S}}, i\rangle \otimes |\epsilon_{\mathrm{B}} , j\rangle
\langle \epsilon_{\mathrm{S}}, i|\otimes\langle \epsilon_{\mathrm{B}}, j|
\end{equation}
where the $|\epsilon_{\mathrm{S}}, i\rangle$ ($i=1, \dots, n_{\mathrm{S}}(\epsilon_{\mathrm{S}})$) are a basis of energy eigenstates of S with energy $\epsilon_{\mathrm{S}}$, and similarly with S replaced by B; the sum over $i$ goes from $1$ to $n_{\mathrm{S}}(\epsilon_{\mathrm{S}})$, the sum over $j$ goes from $1$ to $n_{\mathrm{B}}(\epsilon_{\mathrm{B}})$ and the sums over $\epsilon_{\mathrm{S}}$ and $\epsilon_{\mathrm{B}}$ are over values which are positive-integer multiples of $\Delta$ and are constrained to have $\epsilon_{\mathrm{S}} + \epsilon_{\mathrm{B}} = E$.
We note that $M$ is related to $n_{\mathrm S}(\epsilon)$ and $n_{\mathrm B}(\epsilon)$ by 
\begin{equation}
\label{sumnorm}
M=\sum_{\epsilon=\Delta}^E n_{\mathrm{S}}(\epsilon)n_{\mathrm{B}}(E-\epsilon)
\end{equation}
where, again, the sum is over energies which are integral multiples of $\Delta$ and we note that, roughly, $M$ will scale with $\Delta$.
It is standard and straightforward to see from (\ref{microsysbath}) that the reduced density operator, $\rho^{\mathrm{microc}}_{\mathrm{S}}$, of S is given by (dropping the `S' subscript on $\epsilon_{\mathrm S}$ from now on)
\begin{equation}
\label{microreduced}
\rho^{\mathrm{microc}}_{\mathrm{S}}=M^{-1}\sum_{\epsilon=\Delta}^E
n_{\mathrm{B}}(E-\epsilon) \sum_{i=1}^{n_{\mathrm{S}}(\epsilon)}
|\epsilon, i\rangle\langle \epsilon, i|.
\end{equation}
The main new development in \cite{Kaythermality} is that we argue that, if in place of $\rho_{\mathrm{microc}}$,
we choose at random a pure totem state vector $\Psi$ from the subspace ${\cal H}_M$, then, for a very wide range of system and energy-bath densities of states, the reduced density operator of $|\Psi\rangle\langle\Psi|$ on the system S will, with a high probability, be close to a density operator of the form $\rho^{\mathrm{modapprox}}_{\mathrm{S}}$ given by
\begin{equation}
\label{purereduced}
\rho^{\mathrm{modapprox}}_{\mathrm{S}}=M^{-1}\left(\sum_{\epsilon=\Delta}^{E_c}
n_{\mathrm{B}}(E-\epsilon)\sum_{i=1}^{n_{\mathrm{S}}(\epsilon)}
 |\epsilon, i\rangle\langle \epsilon, i|+
\sum_{\epsilon=E_c+\Delta}^E
n_{\mathrm{S}}(\epsilon)\sum_{i=1}^{n_{\mathrm{B}}(E-\epsilon)}
|\widetilde{\epsilon, i}\rangle\langle
\widetilde{\epsilon, i}| \right )
\end{equation}
where we define $E_c$ to be the energy value at which 
$\sigma_{\mathrm{S}}(E_c)=\sigma_{\mathrm{B}}(E-E_c)$. When $\epsilon > E_c$, 
the $|\widetilde{\epsilon, i}\rangle$ in this formula denote the elements of an orthonormal basis of an $n_{\mathrm{B}}(E-\epsilon)$-dimensional subspace of the ($n_{\mathrm{S}}(\epsilon)$-dimensional) energy-$\epsilon$ subspace of ${\cal H}_{\mathrm{S}}$ which will depend on $\Psi$.  The formula (\ref{purereduced}) is useful since the mean energy of the system ($n=1$ in the next formula) and other $n$th moments of the system energy when the totem has the state vector $\Psi$ will, with high probability, be independent of $\Psi$ and be given as ${\mathrm {tr}}(\rho^{\mathrm{modapprox}}_{\mathrm{S}}H_{\mathrm S}^n)$  and also the S-B entanglement entropy (equal \cite{Kay1, Kaythermality} to the von Neumann entropy of the reduced density operator of the system, S, and also equal to the von Neumann entropy of the reduced density operator of the energy bath, B) will, with high probability, be largely independent of $\Psi$ and close to the von Neumann entropy $S^{\mathrm{modapprox}}_{\mathrm{S}}=-k\,{\mathrm {tr}}(\rho^{\mathrm{modapprox}}_{\mathrm{S}}\log(\rho^{\mathrm{modapprox}}_{\mathrm{S}}))$ (which is automatically equal to $S^{\mathrm{modapprox}}_{\mathrm{B}}=-k\,{\mathrm {tr}}(\rho^{\mathrm{modapprox}}_{\mathrm{B}}\log(\rho^{\mathrm{modapprox}}_{\mathrm{B}})$); to calculate the values of any of these quantities we do not need to know which subspaces those $n_{\mathrm{B}}(E-\epsilon)$-dimensional subspaces are, nor to know how they depend on $\Psi$.  In particular, one easily derives from (\ref{purereduced}) that 
\begin{equation}
\label{contpureentropy}
S^{\mathrm{modapprox}}_{\mathrm{S}}=S^{\mathrm{modapprox}}_{\mathrm{B}}=
k\int_0^{E_c} P_{\mathrm{S}}(\epsilon)\log\left (
\frac{\sigma_{\mathrm{S}}(\epsilon)}{P_{\mathrm{S}}(\epsilon)}\right)d\epsilon+k\int_{E_c}^E P_{\mathrm{S}}(\epsilon)\log\left (
\frac{\sigma_{\mathrm{B}}(E-\epsilon)}{P_{\mathrm{S}}(\epsilon)}\right)d\epsilon
\end{equation}
where $P_{\mathrm{S}}(\epsilon)$ denotes the {\it energy probability density} of S.
\begin{equation}
\label{enprobdens}
P_{\mathrm{S}}(\epsilon)=
\frac{\Delta}{M}\sigma_{\mathrm{S}}(\epsilon)\sigma_{\mathrm{B}}(E-\epsilon)
= \frac{1}{M\Delta}n_{\mathrm{S}}(\epsilon)n_{\mathrm{B}}(E-\epsilon).
\end{equation}
(We note that $P_{\mathrm{B}}(\epsilon)=P_{\mathrm{B}}(E-\epsilon)$ and the mean energy [and other moments of energy] of S and B arise, on both traditional and modern approaches, as integrals of [powers of] $\epsilon$ times $P_{\mathrm{S/B}}(\epsilon)$.) 
Equation (\ref{contpureentropy}) is to be contrasted with the formula,
\begin{equation}
\label{contmicroentropy}
S^{\mathrm{microc}}_{\mathrm{S}}=
k\int_0^E P_{\mathrm{S}}(\epsilon)\log\left (
\frac{\sigma_{\mathrm{S}}(\epsilon)}{P_{\mathrm{S}}(\epsilon)}\right)d\epsilon,
\end{equation}
for the von Neumann entropy, $S^{\mathrm{microc}}_{\mathrm{S}}$, of $\rho^{\mathrm{microc}}_{\mathrm{S}}$ (and a similar formula for the, now, in general, different, quantity $S^{\mathrm{modapprox}}_{\mathrm{B}}$).

(Strictly \cite{Kaythermality} we should replace $\int_0^E d\epsilon$ in (\ref{contpureentropy}) and (\ref{contmicroentropy}) by $\sum_{\epsilon=\Delta}^E$ where the sum is over integer multiples of $\Delta$.)

Recalling that $M$ scales with $\Delta$, we see from the first equality of 
(\ref{enprobdens}) that the energy probability density is independent of $\Delta$ and therefore, from 
(\ref{contpureentropy}) and (\ref{contmicroentropy}) , we draw the important conclusion that neither $S^{\mathrm{modapprox}}_{\mathrm{S}}$, nor $S^{\mathrm{microc}}_{\mathrm{S}}$, depend on $\Delta$.

There are standard arguments that, when the system, S, is small (i.e.\ when the density of states of S is much `less dense' than the density of states of B) the $\rho^{\mathrm{microc}}_{\mathrm{S}}$ of (\ref{microreduced}) will be close to the density operator of a Gibbs state at inverse temperature $k\,d\log\sigma_{\mathrm B}(E)/dE$.  One of the uses of our new formula
(\ref{contpureentropy}) is that it enables us to see straight away that the same must be true, with high probability, when the totem state vector is a random pure state vector in ${\cal H}_M$, since, when the system is small, $E_c$ will be close to $E$ and the formulae (\ref{microreduced}) and (\ref{purereduced}) will coincide.  Thus we easily re-obtain the `typicality' result of \cite{GoldsteinLebowitzetal}.  

Our main interest in the formalism outlined above is so as to attempt to model a black hole equilibrium state by making the identifications: S $\leftrightarrow$ matter; B $\leftrightarrow$ gravity.  We note first that, with most choices for the density of states functions, $\sigma_{\mathrm S}$ and 
$\sigma_{\mathrm B}$, once one drops the assumption that S is small, it will no longer be true any more, in either of the traditional or modern approaches, that S (or B) will be thermal (as we illustrate in detail for power-law densities of states in \cite{Kaythermality}).  However, they will be thermal in certain well-defined approximate senses if, with either the traditional or modern approaches, both the densities of states are taken to have the quadratic exponential form $\sigma_{\mathrm{S}/\mathrm{B}}
=K\exp(q\epsilon^2)$ (with inverse temperature, $1/T= 2kqE$) or if they are taken to have the exponential form $\sigma_{\mathrm{S}/\mathrm{B}} =c\exp(b\epsilon)$ (with inverse temperature $1/T=kb$).  In the quadratic exponential case, we find that (up to small order 1 corrections) $S^{\mathrm{microc}}_{\mathrm{S}}=kqE^2/2$ while $S^{\mathrm{modapprox}}_{\mathrm{S}}=0$ .  In the exponential case, we find \cite{Kaythermality} that (up to small logarithmic terms)
\begin{equation}
\label{expentropy}
S^{\mathrm{microc}}_{\mathrm{S}}=kbE/2, \ \ \hbox{while} \ \ S^{\mathrm{modapprox}}_{\mathrm{S}}=kbE/4.
\end{equation}

The above results for equal quadratic exponential densities of states may seem, at first sight, to be good news for the possibility of a traditional microcanonical model of black hole equilibrium states and it is, indeed, intriguing that, if we identify $q$ with  $2\pi G$ and equate the mean energy, $E/2$, of B with the black hole mass, $\cal M$, then the above formula $1/T=2kqE$ agrees with the Hawking inverse temperature $1/T=8\pi kG{\cal M}$ for a Schwarzschild black hole and, at the same time, the formula $S^{\mathrm{microc}}_{\mathrm{S}}=kqE^2/2$ for the entropy of B agrees with the Hawking entropy formula $S=4\pi kG{\cal M}^2$ for the same black hole.  And the fact that $S^{\mathrm{modapprox}}_{\mathrm{S}}=0$ might seem to be bad news for any modern model and hence also bad news for our matter-gravity entanglement hypothesis.  However, as we argue at greater length in \cite{Kaythermality}, such conclusions are unwarranted since we do not expect either the traditional microcanonical or our modern formalism to be applicable because they assume weak coupling, whereas, if the relevant degrees of freedom are assumed to be those of the gravitational field itself, black holes are intrinsically strong-coupling objects.

Where weak-coupling scenarios do become relevant is with the string-theory approach to quantum gravity, according to which, as first argued by Susskind \cite{Susskind}, as one decreases the string coupling constant, black holes go over to certain states of string theory which we know how to count.  This has led to calculations (an important early paper was that of Strominger-Vafa \cite{StromingerVafa})  which, in the case of certain extremal and near-extremal black holes, exactly re-obtain the Hawking entropy formula $S=kA/4G$ as $k$ times the logarithm of the degeneracy of the string-theoretic energy level which goes over to the relevant black hole at strong string coupling.  Moreover, for far-from-extremal black holes, such as Schwarzschild, Horowitz and Polchinski \cite{HorowitzPolchinski}, building directly on \cite{Susskind}, have obtained similar, but now semi-qualitative, agreement by arguing that, as one scales the string length scale, $\ell$, up and the string coupling constant, $g$, down from their physical values, keeping $G=g^2\ell^2$ fixed, a (4-dimensional, Schwarzschild) black hole of mass $\cal M$ will go over to a long string with roughly the same energy, $\epsilon={\cal M}$.  The density of states of such a long string, for small enough $g$, is known, very roughly (i.e.\ omitting an inverse-power prefactor) to take the exponential form, $\sigma_{\mathrm{ls}}(\epsilon) = C_{\mathrm{ls}}e^{\ell \epsilon}$ ($C_{\mathrm{ls}}$ a constant with the dimensions of inverse energy of the same order of magnitude as $\ell$).  \cite{Susskind, HorowitzPolchinski} then point out that the `logarithm' of this is approximately given by $S_{\mathrm{ls}}=\ell\epsilon$ and propose that this should be equated with the entropy of a (Schwarzschild) black hole provided that one does the equating when, to within an order of magnitude or so, $\ell=G{\cal M}$. Combining these latter two equations (and replacing $\epsilon$ by ${\cal M}$) one arrives at the prediction that the entropy of the black hole will be a moderately sized constant times $kG{\cal M}^2$ which agrees, up to an undetermined value for the constant, with the Hawking value, $4\pi kG{\cal M}^2$, for the entropy of a black hole.                                                                                                                                                                                                                                                                                                                                                                                                                                           

It is often argued that the successful agreement of the results of the Strominger-Vafa and the Susskind-Horowitz-Polchinski calculations with the Hawking black hole entropy formulae constitute `understandings' of black hole entropy and, further, in view of the fact that string theory is a conventional quantum theory with a unitary dynamics, that the Information Loss Puzzle is thereby resolved.  However, the calculations in, say, \cite{StromingerVafa} are actually not of {\it entropies} but of {\it degeneracies}, and it remains a challenge to explain why these degeneracies agree with black hole entropies.  After all, the degeneracy of the $n$th energy level of the textbook Hydrogen atom Hamiltonian is $n^2$ but we would not conclude that the Hydrogen atom has an entropy of $k\,\log n^2$!  The difficulty is compounded in the work of \cite{Susskind, HorowitzPolchinski} since that is couched in terms of a density of states, $\sigma_{\mathrm{ls}}$, rather than a degeneracy and, one should of course really multiply this by a constant with the dimensions of energy before one can take its logarithm and no such constant is supplied.  Of course, the logarithm of the degeneracy of an energy level is the same thing as the entropy of the microcanonical state for that energy level.  But if one were to argue that the string theory state which goes over to the Stominger Vafa black hole was just this microcanonical state, then the Information Loss Puzzle would surely arise again:  How, in a dynamical process of black hole formation, can a pure string theory state evolve into such a (non-pure) microcanonical state?

In view of these difficulties, we would rather conclude from the undoubted success of these calculations that they are important hints towards a full understanding of black hole entropy and towards a resolution to the Information Loss Puzzle but they cannot be the final word on these questions.

What we wish to argue here is that our matter-gravity entanglement hypothesis (which, as we discussed above, incorporates a solution to the Information Loss Puzzle) may take us a step closer towards such a final word:  In particular, we propose that the Horowitz-Polchinski scenario be replaced by a scenario in which,  as one scales the string length scale, $\ell$, up and the string coupling constant, $g$, down from their physical values, keeping $G=g^2\ell^2$ fixed, an {\it equilibrium state} consisting of a  (4-dimensional) Schwarzschild black hole of mass $\cal M$ in contact with its (mostly matter) atmosphere in a box of given total energy, $E$, will go over to an equilibrium state of similar total energy, $E$, consisting of a long string, with mean energy, $\bar\epsilon$, roughly equal to ${\cal M}$, in contact with an atmosphere of small strings in a, suitably rescaled, box.  If we ignore certain inverse-power prefactors, each of the long string and the stringy atmosphere densities of states, $\sigma_{\mathrm{ls}}$ and, say, $\sigma_{\mathrm{sa}}$, will take the form $Ce^{\epsilon\ell}$ (with different values, $C_{\mathrm{ls}}$ and, say, $C_{\mathrm{sa}}$, of $C$ but this is unimportant \cite{Kaythermality}).  If we now adopt the (rough) correspondences (see the relevant Endnote in \cite{Kayprefactor} for further discussion):  long string $\leftrightarrow$ gravity ($\leftrightarrow$ B), stringy atmosphere $\leftrightarrow$ matter ($\leftrightarrow$ S), then our above formalism and the results (\ref{purereduced}) and (\ref{contpureentropy}) are applicable, with these densities of states in place of $\sigma_{\mathrm{B}}$ and $\sigma_{\mathrm{S}}$:  In the modern scenario with a pure state vector, $\Psi$, chosen at random from the state space, ${\cal H}_M$, the mean energy of the long string (as also of the stringy atmosphere) will be $E/2$ and the entropy (i.e.\ the S-B entanglement entropy) will, by the second equation in (\ref{expentropy}) with $b=\ell$, be $k\ell E/4$.  Assuming we can equate this mean energy with $\cal M$ and this entropy with the entropy of the black hole equilibrium state when $\ell$ is, say, $XG{\cal M}$, we predict an entropy for the black hole equilibrium of $kXG{\cal M}^2/2$ while it will be approximately thermal with inverse temperature, $1/T=k\ell=kXG{\cal M}$.  It is intriguing to note that there is a single value of $X$, namely $X=8\pi$, for which these agree simultaneously with the Hawking entropy, $4\pi kG{\cal M}^2$, and the Hawking inverse temperature, $8\pi kG{\cal M}$.  And, in view of the factor of 2 discrepancy between the microcanonical and modern entropies in (\ref{expentropy}), it is clearly not possible to obtain such simultaneous agreement of temperature and entropy with the, non-modern, microcanonical scenario.   However we caution that (as we discuss further in \cite{Kayprefactor}) it is not clear whether this feature of our modern model with exponential densities of states survives when we improve the model to include the appropriate prefactors mentioned below.  Nevertheless, our main point is that a `modern' model for black hole entropy, based on our matter-gravity entanglement hypothesis seems able to predict a temperature of the order of the Hawking temperature and an entropy of the order of the Hawking entropy.

In further work, on this proposed new `modern' string-theoretic understanding of black hole entropy, we have studied a more realistic model which includes the appropriate approximately (i.e.\ for non-small energies) inverse-power prefactors multiplying the exponentials in the density of states of a single string.   These prefactors are in fact crucial and justify our assumption above that the string equilibrium in a box will consist of one long string with a stringy atmosphere of small strings.  Our results, which are outlined separately \cite{Kayprefactor}, appear to confirm the broad features of the result obtained here, although there are also significant and interesting (and encouraging) differences.

\end{document}